\documentclass[12pt]{iopart}

\usepackage{graphicx}

\begin{document}

\title{Multiplicity distributions in rapidity for Pb-Pb and p-Pb central collisions from a simple model}

\author{S. Abreu, J. Dias de Deus and \underline{J.G. Milhano}}

\address{Instituto Superior T\'ecnico/CENTRA, Av. Rovisco Pais, P-1049-001 Lisboa, Portugal}



\vspace{0.5cm}

The simple model \cite{DiasdeDeus:2007wb}  for the distribution of rapidity extended objects (longitudinal glasma colour fields or coloured strings) created in a heavy ion collision combines the generation of lower centre of mass rapidity objects from higher rapidity ones with asymptotic saturation in the form of the well known logistic equation for population dynamics 
\begin{equation}\label{eq:log}
	\frac{\partial \rho}{\partial (-\Delta)} = \frac{1}{\delta} (\rho - A \rho^2)\, ,
\end{equation}
where $\rho \equiv \rho (\Delta, Y)$ is the particle density, $Y$ is the beam rapidity, and $\Delta \equiv |y| - Y$. The $Y$-dependent limiting value of $\rho$ is determined by the saturation condition  $\partial_{(-\Delta)} \rho = 0 \longrightarrow\rho_Y = 1/A$, while the separation between the low density (positive curvature) and high density (negative curvature) regions is given by ${\partial^2}_{(-\Delta)} \rho\big |_{\Delta_0} = 0\longrightarrow \rho_0 \equiv \rho( \Delta_0, Y) =  {\rho_Y}/{2}$. Integrating (\ref{eq:log}) we get
\begin{equation}\label{eq:rhoint}
	\rho( \Delta, Y) = \frac{\rho_Y}{e^{\frac{\Delta - \Delta_0}{\delta}}+1}\, .
\end{equation}
In the String Percolation Model \cite{DiasdeDeus:2007wb} the particle density is proportional, once the colour reduction factor is taken into account, to the average number of participants $\rho\propto N_A$; the normalized particle density at mid-rapidity is related to the gluon distribution at small Bjorken-$x$, $\rho\propto \exp^{\lambda\, Y}$; and the dense-dilute separation scale decreases, from energy conservation, linearly with $Y$,  $\Delta_0=-\alpha Y$ with $0<\alpha <1$. Rewriting (\ref{eq:rhoint}) in rapidity
\begin{equation}\label{eq:rhoy}
	\rho \equiv  \frac{dN}{dy} =  \frac{N_A \cdot e^{\lambda Y}}{e^{\frac{|y|- (1-\alpha) Y}{\delta}} +1}\, .
\end{equation}
The values $\lambda=0.247$, $\alpha=0.269$ and $\delta=0.67$ for the parameters in the solution (\ref{eq:rhoy}) are fixed by an overall fit \cite{Brogueira:2006nz} of  $Au-Au$ RHIC data \cite{Back:2004je}.

\begin{figure}[h] 
   \centering
   \includegraphics[angle=0,width=10cm]{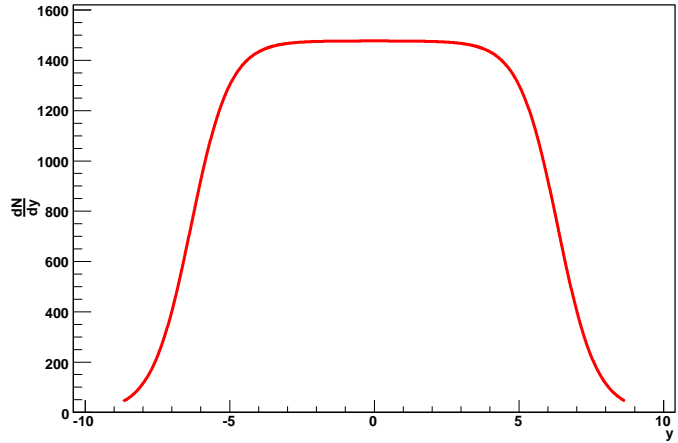} 
   \caption{$\frac{dN}{dy}$ from (\ref{eq:rhoy}) for $Pb-Pb$ (0-10\% central) collisions at $\sqrt{s}=5.5$ TeV with $N_A=173.3$}
   \label{fig:PbPb}
\end{figure}

In Figure \ref{fig:PbPb} we show the predicted multiplicity distribution for the 10\% most central $Pb-Pb$ collisions at $\sqrt{s}=5.5$ TeV with $N_A=173.3$ taken from the Glauber calculation in \cite{Kharzeev:2004if}.

\begin{figure}[h] 
   \centering
   \includegraphics[angle=0,width=10cm]{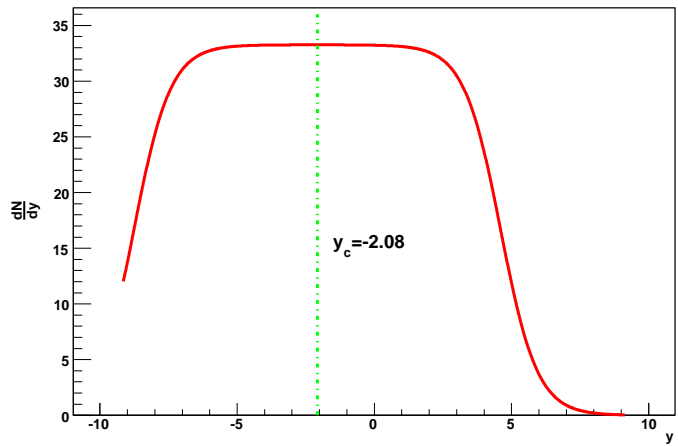} 
   \caption{$\frac{dN}{dy}$ from asymmetric version of (\ref{eq:rhoy}) \cite{DiasdeDeus:2007wb} for $p-Pb$ (0-20\% central) collisions at $\sqrt{s}=8.8$ TeV with $N_{part}=13.07$.}
   \label{fig:pPb}
\end{figure}

In  Figure \ref{fig:pPb} we show the predicted multiplicity distribution for the 20\% most central $p-Pb$ collisions at $\sqrt{s}=8.8$ TeV with $N_{part}=13.07$ also from \cite{Kharzeev:2004if}. In this case the solution  (\ref{eq:rhoy}) have been modified to account for the asymmetric geometry and the shift of the centre of mass of the system relatively to the laboratory centre of mass \cite{DiasdeDeus:2007wb}. The resulting rapidity shift $y_c = -2.08$ is marked in the figure.

J.~G.~M.  acknowledges the financial support of  the Funda\c c\~ao para a Ci\^encia e a Tecnologia  of Portugal (contract SFRH/BPD/12112/2003).

\end{document}